\documentclass[prd]{revtex4}

\newcommand{\be}{\begin{equation}}
\newcommand{\ee}{\end{equation}}
\newcommand{\ba}{\begin{eqnarray}}
\newcommand{\ea}{\end{eqnarray}}

\begin{document}

\title{Emergent Photons and Gravitons:  The Problem of Vacuum Structure
\footnote{Talk presented at the Fifth Meeting on CPT and Lorentz
Symmetry, University of Indiana, June 28-July 2,
2010.}\footnote{This work was supported in part by Contract No. DE
AC02-76F00515.}}

\author{J. D.\ BJORKEN}

\affiliation{SLAC National Accelerator Laboratory\\
2575 Sand Hill Road\\
Menlo Park, CA  94025/Pacific, USA}
\email{bjorken@slac.stanford.edu}

\begin{abstract}
We discuss vacuum condensates associated with emergent QED and with
torsion, as well as the possible role of the Kodama wave function in
quantum cosmology.
\end{abstract}

\maketitle

\section{Introduction}
It is a great pleasure to be here and participate in this meeting.
The subject matter and the SME (Standard Model Extension) approach
pioneered by our host Alan Kostelecky' has been of great interest to
me for a long time. In this talk, I will skim through a few specific
ideas which are on my mind, and in general eschew the grand
overview.\cite{arxiv} However, I will first define some ground
rules that I shall use. They greatly overlap with Alan's ground
rules, so these introductory remarks should also serve as a very
brief review of Alan's program. The problem addressed in this
meeting is the Big Problem:  what happens to the standard model and
Einstein gravity when the energy scale approaches and exceeds the
Planck scale? I see this problem as quite analogous to the one which
was faced in the 1950's and 1960's: what happens when the energy
scale exceeds the natural scale of the strong interactions? In
hindsight, and with some liberties taken with history, the low
energy theory could be described in terms of an effective action.
The ingredients of this effective action consisted of self-coupled
pions and nucleons, with an apparent isospin symmetry and a slightly
broken chiral symmetry, responsible for the relatively small mass of
the pion relative to the natural strong-interaction scale. On the
high-energy, short-distance side of the strong interaction barrier,
the appropriate degrees of freedom turned out to be quarks and
gluons, with the original degrees of freedom reinterpreted as
non-fundamental, composite fields. For the Big Problem, I presume
that the fate of the pion is not so different than what the fate of
the photon, the gluons, the electroweak gauge bosons, and the
graviton will be--they become described in terms of more fundamental
degrees of freedom. It is quite probable that new concepts, as new
as color and quarks were for the strong interactions, will be
required--especially regarding the structure of spacetime itself. The
challenge is formidable, because experiments will be much scarcer
than they were in breaking through the strong-interaction barrier.
But we should never give up.
\section{Emergent QED and Gravity}
The idea that the photon might be a Goldstone boson of a theory of
spontaneously broken Lorentz covariance goes back a long way. I
myself made a try,\cite{bj} copying closely the Nambu-Jona-Lasinio
formalism\cite{NJL} for the Goldstone pion. This idea is better
expressed nowadays in terms of effective field theory language,
where one can easily presume that the Maxwell Lagrangian one wants
for QED is readily obtained via loop-diagram radiative effects. I
prefer to focus on this effective Lagrangian as much as possible,
setting aside the detailed mechanism of how it arises (at least for
now) as much as possible. So the interesting problem for me is what
other terms in the effective Lagrangian come along with the
ride--there may be gauge and/or Lorentz violating terms as well. The
general attack on this question is beautifully expressed in terms of
Alan's Standard Model Extension (SME). I personally opt to dumb down
the problem by asking which of the violating terms, not present in
textbook QED, might be the most important. My general rule is that
gauge-variant terms should be suppressed, probably by very small
coefficients, and that Lorentz-violating terms are even more
suppressed. This is, because of the SME constraints, a matter of
simple survival.

A few years ago, I revisited the subject,\cite{jdb} and guessed that
the leading gauge-variant term might be a Mexican hat potential,
with a huge, GUT-scale vacuum expectation value $M$ for the gauge
potential. I chose the quartic coupling constant to be extremely
small, of order $10^{-30}$, in such a way that it would vanish in
the limit of vanishing dark energy. \be
L=\frac{1}{2}(\textbf{E}^{2}-\mathbf{B}^{2})-\frac{\mu}{M}(\phi^{2}-\textbf{A}^{2}-M^{2})
- (\rho \,\phi -\textbf{J}\cdot\textbf{A}) \ee Following the
procedure used in the effective theory of Goldstone pions, i.e.
eliminating the sigma meson in terms of the three pion fields, one
can eliminate the electrostatic potential $\phi$ in terms of the
vector potential $\textbf{A}$, remaining at the bottom of the
Mexican hat potential well. Implementing this leaves behind only the
Maxwell term in the Lagrangian, but with a nonlinear relation
between electric field $\textbf{E}$ and gauge potential
$\textbf{A}$. \be \phi = \sqrt{\textbf{A}^{2} + M^{2}} = M +
\frac{\textbf{A}^{2}}{2M} + \cdot\cdot\cdot
 \ee \be
\textbf{E} = - \dot{\textbf{A}} - \nabla (\frac{\textbf{A}^{2}}{2M})
+ \cdot\cdot\cdot
 \ee When the dust settles, at
tree level the effect of adding the extra Mexican-hat term amounted
to fixing the gauge--although the gauge one gets is a curiously
nonlinear one. The Gauss-law Maxwell equation of constraint (in the
presence of an external conserved charge-density $\rho$ becomes  \be
\frac{\partial\Gamma}{\partial t} = \nabla \cdot (\mathbf{v}\Gamma)
\ee \be \Gamma = \nabla \cdot \textbf{E} - \rho \quad\quad
\mathbf{v} = \frac{\textbf{A}}{\sqrt{\textbf{A}^{2}+ M^{2}}} \ee If
gauge invariance is broken, as above, then there is a preferred
gauge, in terms of which the theory most closely follows the
dynamics of the underlying, hidden degrees of freedom. In general,
it makes sense to guess the "most probable gauge". My choice is
temporal gauge, and the above scenario is a specific way to express
this choice. In temporal gauge, the longitudinal-photon degrees of
freedom are, in a sense, dynamical, because they have non-vanishing
canonical momenta. However, in practice the Gauss-law constraint
makes these degrees of freedom act like a Bose condensate, described
by only a few classical degrees of freedom. It is interesting that,
in CPT2001, Nambu discussed just this point in his talk,\cite{nambu}
and ascribed this idea to work of Dirac\cite{pam} in the 1950's.

To me it would be very interesting if somehow this
longitudinal-photon condensate might somehow be activated. So
recently I gave it a try. The game is to stay with the Mexican-hat
picture above, but to assume that the vacuum gauge-potential
condensate has spacetime dependence. A very simple, cosmological
type of behavior is to assume \ba F_{\mu\nu} = 0 && \quad  A_{\mu} =
\partial_\mu\, \Lambda(r,t) \ea

Our previous example set  $ \Lambda = M\,t$. If we choose instead
\be \Lambda = M \, \tau \quad \quad \tau^{2} = t^{2} -
\frac{r^{2}}{c^{2}} \ee it follows that \be \phi = \frac{M t }{\tau}
\quad\quad \textbf{A}= - \frac{M \textbf{r}}{c^{2}\tau} \quad\quad
\phi^{2} - c^{2}\mathbf{A}^{2} = M^{2} \ee This can be constructed
from the same Mexican-hat potential as before, provided that $c =
1$. I put in the Lorentz violation mostly (but not entirely) for
fun, because the solution admits so easily the generalization.

The net result of this construction is a vacuum which will become,
or which has been, unstable, depending upon whether we live in the
past or future "lightcone" associated with the gauge function
$\Lambda$. It seems to me that this might be a mechanism for
catalyzing the cosmological "reheating"  transition, because the
onset of the instability outraces even the accelerated expansion of
the universe.

What about emergent gravitons? The idea goes back to
Sakharov\cite{sakharov},and the Einstein-Hilbert action is arguably
easy to obtain via radiative loops. Again the problem is what else,
if anything, comes along for the ride. A general attack can quickly
lead to quite a mess.\cite{KT} At the opposite extreme, I might
guess that the most important violating term is a potential
$V(\sqrt{g})$, depending only upon the determinant $g$ of the
metric. The Einstein equations are than easy to obtain, and they
will make trouble unless $V'$ vanishes. This leads to a fixed value
of the determinant, and a consequent "emergent unimodular gravity".
So at this level I only see gauge fixing as the output consequence.
Quite a lot more can be said about this approach,\cite{HT} but Alan
is better equipped than I to say it\cite{AKpotting}.

\section{Torsion and CP Violation}

If one wishes to synthesize the standard model with gravity it would
seem that the quark and lepton degrees of freedom have to be a
rather central part of the story. This means that the coupling of
gravity to spinors is a very relevant issue. The standard metric
formalism is ill-suited to this task and one should use the
first-order Einstein-Cartan (Palatini) formalism instead. The ten
degrees of freedom describing the metric tensor are replaced by
forty. Twenty-four are connection variables $\omega$. They are
essentially gauge potentials, a set of six four-vectors in
spacetime, which live in the adjoint representation of an "internal"
O(3,1) gauge group. The other sixteen degrees of freedom are the
tetrad variables $e$, which are a set of four spacetime four-vectors
living in the vector representation of O(3,1). The Yang-Mills
curvature of the connection is called $R$. There are evidently
thirty-six components of that beast. And the usual metric tensor $g$
is defined in terms of a quadratic form built from the tetrad; the
O(3,1) indices are contracted via a Minkowski metric.

In this formalism there are six terms which are in a sense
"leading", at least in descriptions of gravity commonplace in the
loop-gravity community. Three of them are present in ordinary metric
gravity. They are the Einstein-Hilbert term, the
cosmological-constant term, and the Gauss-Bonnet (Euler) topological
term: \ba L_{EH} = e^{A}\wedge e^{B} \wedge R^{CD}
\wedge\epsilon_{ABCD}\nonumber \\ L_{CC} = e^{A} \wedge e^{B} \wedge
e^{C} \wedge e^{D} \epsilon_{ABCD}\nonumber \\ L_{GB} = R^{AB}
\wedge R^{CD} \epsilon_{ABCD} \ea (These indices $ABCD$ live in the
O(3,1) internal space; the wedge products act on spacetime indices.)

Variation of such an action with respect to the connections $\omega$
yields a constraint on the connection, namely that it be the usual
Levi-Civita connection described by the Christoffel symbols.
Variation with respect to the tetrad variables yields the Einstein
equations. The Gauss-Bonnet/Euler term is the divergence of a
topological current and does not contribute to the equations of
motion. But it will reappear later in this story.

In this formalism, the connection does not start out ab initio as
Levi-Civita, as in metric gravity. The difference between a general
connection and the Levi-Civita choice is described by what is called
torsion. While it is very arguable that torsion can be neglected in
the classical theory, it is much less obvious that it can at the
quantum level, especially when spinor degrees of freedom are
included. If torsion is admitted, it is also more natural to admit
additional terms into the gravitational action, in particular terms
which are CP odd. It turns out that there are three "leading terms"
that can most naturally be included. The most interesting one is
called the Holst term,\cite{holst} which is a close analog of the
Einstein-Hilbert term. The other two, the Pontryagin and Nieh-Yan
terms,\cite{CZ} are both topological. Like the Gauss-Bonnet term,
they do not affect equations of motion. But they might lead to
subtle quantum effects similar to what happens with the (Pontryagin)
topological term $\mathbf{E}\cdot\mathbf{B}$ in QCD: \ba L_{H} =
e^{A} \wedge e^{B} \wedge R_{AB} \nonumber\\ L_{P} = R^{AB} \wedge
R_{AB} \nonumber\\L_{NY} + L_{H} = (De)^{A} \wedge (De)_{A} \ea
(Here $D$ is the covariant derivative with respect to the connection
$\omega$; see Equation 13 below for some details.)

I find it impressive that this formalism for pure gravity so
naturally admits CP violation. And from an effective-field-theory
point of view, there is no reason to omit such terms ab initio,
because we know that CP violation occurs in nature.

It is especially important to understand the role of the Holst term,
which is not pure topological. The coefficient of the Holst term in
the action is inversely proportional to what is known as\cite{gamma}
the Barbero-Immirzi parameter $\gamma$. In the first order
formalism, especially as elucidated by the loop-gravity community,
this Holst term creates mixing between torsion degrees of freedom
and ordinary metric degrees of freedom, but in such a way that the
Einstein equations for ordinary macroscopic applications remain
unaffected. But the formal canonical theory, used to set up
quantization, is deeply affected.\cite{bojowald}

\section{An Axial-Vector Condensate}
Just as I did for the Goldstone photon, I have tried to "activate"
such torsion degrees of freedom in as simple a way as possible, in
order to see how they might enter into phenomenology. This led to
presuming that there might exist, for some fermionic degrees of
freedom (either standard-model or beyond-the-standard-model), a
Lorentz- violating vacuum condensate of axial vector current: \be
<\overline{\Psi}\gamma_{5}\gamma_{\mu}\Psi> \,\,= \,\,\eta_{\mu}
\rho_{A}
 \ee Here $\eta_\mu$ is a unit timelike vector, at
rest in the CMB rest frame. The torsion part of the connection
appears in the Dirac action. In first order form the Dirac operator
is rewritten as follows: \be \gamma^{\mu}\nabla_{\mu}
\longrightarrow \epsilon_{ABCD}(e^{A}\wedge e^{B}\wedge e^{C}
\gamma^{D})\wedge D \ee The covariant derivative is \be D_{\mu} =
\partial_{\mu} + \frac{1}{2}\omega_{\mu}^{EF}\gamma_{E}\gamma_{F}
 \ee The condensate therefore leads to a source term which couples
linearly to the connection and in particular to a piece of the
torsion known in the trade as contorsion. In the case of
$\Lambda$CDM cosmology the mathematics is simple enough for even me
to do. The nonvanishing components of the tetrad and the connection
are \ba e_{t}^{0} = N \nonumber\\  e_{x}^{1} = e_{y}^{2} = e_{z}^{3} = a(t) \nonumber\\
\omega_{x}^{01} = \omega_{y}^{02} = \omega_{z}^{03} = K(t) \nonumber\\
\omega_{x}^{23} = \omega_{y}^{31} = \omega_{z}^{12} = C(t)
 \ea $N$ is what is known in the trade as lapse, a constant which
determines the rate of ticking of the FRW clock. It can be set to
unity after the equations of motion have been obtained from the
variational principle. $a(t)$ is the familiar FRW scale factor. The
quantity $K(t)$ is known in the trade as extrinsic curvature, while
$C(t)$ measures the contorsion.

With this starting point, it is not too hard to explicitly construct
equations of motion, solve them, and connect this language to the
usual textbook metric-gravity language. When the dust settles, one
finds that there is a dark-energy-like contorsion given by \be C(t)
= \frac{2\pi a(t) \gamma^{2} \rho_{A}}{M_{\mathrm{pl}}^{2}
(1+\gamma^{2})} \ee ($\gamma$ is the Barbero-Immirzi parameter.)

There is an important combination $A(t)$ of extrinsic curvature and
contorsion \be A(t) = K(t) + \frac{C(t)}{\gamma}
 \ee The Einstein equations, expressed in terms of $A$, are
essentially unaffected by the presence of the nonvanishing
contortion $C$. Its only effect is to renormalize the dark energy
scale. Let $H$ be the asymptotic, dark-energy-dominated, expansion
rate, as given by $a(t) = e^{Ht}$. This quantity depends upon $C$,
such that \be H^{2} = H_{\mathrm{cc}}^{2} -
\frac{(4\pi\gamma\rho_{A})^{2}}{M_{\mathrm{pl}}^{4} (
1+\gamma^{2})}.
 \ee Here $H_{\mathrm{cc}}$ is evidently the value the Hubble parameter
would take in the absence of torsion and the axial condensate. If
this renormalization of the dark energy scale is of order unity, one
has \be \frac{4\pi\gamma\rho_{A}}{\sqrt{1+\gamma^{2}}}\,\,\sim
\,\,HM_{\mathrm{pl}}^{2}\,\,\sim\,\,10^{-60}M_{\mathrm{pl}}^{3}\,\,\sim\,\,(10^{-20}M_{\mathrm{pl}})^{3}\,\,
\sim\,\,\Lambda_{QCD}^{3} \ee This is what I call the Zeldovich
relation: in natural units the cube of the QCD scale is of order the
Hubble scale. It was noticed by Zeldovich\cite{zeld} in 1967 and has
been occasionally been rediscovered in the interim.\cite{carniero} I
encounter it often in my speculative excursions into trying to
understand the dark energy problem, and I now take it seriously. I
find that this is a minority viewpoint. Most people seem to dismiss
the Zeldovich relation as a numerical coincidence.

This axial condensate has another consequence. Because all spinor
degrees of freedom couple to gravity, they must all feel the effect
of the vacuum contorsion. This leads to a Lorentz-violating term in
the effective action, one which is prominent in the SME
catalog.\cite{AKetal} \be L' =
b_{\mu}\overline{\psi}\gamma^{\mu}\gamma_{5}\psi \ee The condensate
contribution to this Kostelecky' $b$ - parameter is \be b_{\mu} =
\eta_{\mu} \frac{2\pi\gamma^{2}\rho_{A}}{M_{\mathrm{pl}}^{2}( 1 +
\gamma^{2})} \ee Here $\eta_{\mu}$ is a unit timelike vector, pure
timelike in the CMB rest frame. If the Zeldovich relation holds,
then \be b_{\mu} \leq 10^{-33} eV \ee The effect is a billion times
smaller than the experimental limit, unless the condensate density
is taken to be much higher than its "natural" value. Such behavior
would be appropriate for scenarios in which there is a fine-tuned
cancellation of the torsion contribution with a much larger
"primordial" dark energy. I think it is well worth some effort to
push the experimental limits on $b$ if it is not too difficult to do
so.

At this meeting I learned of closely related work of
Poplawski.\cite{nikodem} He uses the QCD quark vacuum condensates
instead of a Lorentz-violating axial condensate to arrive at a very
similar endpoint. This very interesting work is evidently much more
conservative in nature than my utilization of a Lorentz-violating
condensate; hence it is more credible.

\section{Vacuum Phase Density}
The last topic in this potpourri is one in which the Zeldovich
relation again appears. Consider a finite box of spatially flat FRW
$\Lambda $CDM universe, with periodic boundary
conditions\cite{levin} applied ("compactification on a torus"). As
time goes on, this box will expand. The dimensions of the box are
controlled by the FRW scale factor, which evolves according to the
Einstein equations of cosmology. If the box contains only pure dark
energy, it will expand exponentially, with a doubling time for its
volume of about 3 1/2 billion years. As far as I am concerned, all
one needs at initial times is a liter of the stuff; the problem of
what is going on at the microscopic level within such a box is the
fundamental problem of dark energy. The semiclassical wave function
of this box of dark energy is the exponential of a phase factor,
given by the classical action. It turns out to be proportional to
the volume of the box. The coefficient of this phase factor is
linear in the Hubble parameter, in natural units. This leads to the
conclusion that the characteristic volume, for which the "phase
density" is of order $2\pi$, is of order the QCD scale-the Zeldovich
relation again applies.

There is an interesting subplot to this story, which originates in a
variant of first-order gravity invented\cite{MandM} by MacDowell and
Mansouri and elaborated recently by Freidel and
Starodubtsev.\cite{FS} The idea is to synthesize the tetrad and
connection variables ($e$,\,$\omega $) of the first order theory
into a single grand connection $A$ which lives in an internal O(4,1)
space. There are $4 \times 10 = 16 + 24$ slots to fill for such a
connection $A$, just the right number: \be A_{\mu}^{5A} =
H_{\mathrm{cc}}\,\,e_{\mu}^{A}  \quad\quad A_{\mu}^{AB} =
\omega_{\mu}^{AB}\quad\quad (A,B = 0,1,2,3) \ee A 6 x 10 field
strength (curvature) $F$ can be defined in the usual way from the
connection $A$. In terms of this $F$, the Einstein-Hilbert action,
complete with cosmological term and Euler (Gauss-Bonnet) term, takes
a very simple form. First contract the gauge potential (and field
strength) into gamma matrices: \be A_{\mu} \longrightarrow
\frac{1}{2}A_{\mu}^{AB}\gamma_{A}\gamma_{B} \ee Then the Lagrangian
density is simply \be L =
\frac{M_{\mathrm{pl}}^{2}}{H_{\mathrm{cc}}^{2}}\,\mathrm{Tr}\,\gamma_{5}
F \wedge F \ee
 Noteworthy is the large dimensionless coefficient of the action, of
 order $10^{120}$. To get the additional three CP violating terms
 defined in Eqn. 10, one simply makes the replacement $F\rightarrow
 e^{\gamma_{5}\theta} F$ in the above equation; the (constant)
 Barbero-Immirzi parameter $\gamma$ is then given by $\cot\theta$.

This way of expressing gravity is provocative and certainly invites
its use as a starting point for enlarging the theory in some way to
encompass standard model degrees of freedom.\cite{lisi} However,
that is not the issue here. Instead it is easy to find that imposing
a "gauge condition" $F = 0$ for the field strength associated with
the connection $A$ leads to nontrivial solutions. In particular,
deSitter space, which describes our expanding box of dark energy, is
such a solution. According to this interpretation, the vacuum phase
density, given by the exponential of the MacDowell-Mansouri action
(which is quadratic in the field strength $F$) should vanish.

The resolution of this paradox is that the Gauss-Bonnet term,
although pure topological, does contribute vacuum phase. And the
MacDowell-Mansouri construction guarantees that this topological
contribution to the phase density cancels out the contribution given
by the standard metric theory. This Gauss-Bonnet term, complete with
a remarkably large coefficient of $10^{120}$, is essentially what is
known in the loop-gravity community as the Kodama wave
function.\cite{Kodama} However, there it plays a different--and
controversial--role.\cite{Jackiw}

In any case, what is suggested here is that vacuum topology might be
an important ingredient in the understanding of the Zeldovich
relation, of course assuming--as I always do--that it is more than a
numerical accident.

\section{Summary}

I see the main messages in the above discussion to be as follows:

\begin{enumerate}
\item If the "Planck Barrier" is analogous to the "Confinement
Barrier" faced by the physics of the 1950's and 1060's, the
techniques used successfully then may be relevant again. These
include emphasis on local currents and their operator-product
algebra. There are a huge number of two-point functions to consider
in this regard, as well as a huge number of anomaly equations. I do
not think they have been fully investigated. Topology, especially
vacuum topology, probably should be in foreground. Of course there
may be new gauge degrees of freedom to be identified (analogous to
color for the strong interactions)--this is an old story. And it
should be recognized that the role of flavor symmetry, put front and
center in the old days, turned out to be a misleading one. Isospin
is now regarded as an accidental symmetry, and flavor has yet to
submit to a gauge principle. Maybe that will be true again.
Sometimes I like to envisage a flavorless toy universe with only the
third generation included. It is much simpler and cleaner than real
life, so much so that I am occasionally tempted to regard it as a
realistic template for the full theory, to be later embellished by
first- and second-generation infrared decorations.
\item In order to synthesize the standard model degrees of freedom
with gravity, the first order Palatini formalism should be embraced.
It admits torsion and CP violation in a natural way, and this
difference may well end up being real physics, not mere formalism.
Furthermore, the MacDowell-Mansouri / Freidel-Starodubtsev extension
in a sense makes the theory "almost topological", and this viewpoint
may also be a novel one relative to the standard metric language.
\item If gauge invariance is emergent, then there is a preferred
choice of gauge, in terms of which the physics of the underlying
theory is best expressed. And those gauge degrees of freedom, while
largely hidden from view because of the strong SME experimental
constraints, might still be "activated" at some level. I would guess
that the effects of such activation vanish in the limit of vanishing
dark energy. This hypothesis puts the setting of this problem at
only the deepest level, as well as providing some protection from
the often severe SME experimental limits.
\item The cosmological, QCD, and electroweak vacuua, which are all too
easy to compartmentalize because of the disparate distance scales
associated with each, deserve to be synthesized. They occupy the
same region of spacetime, have the same energy, and are coupled
dynamically to each other. The Zeldovich relation may be evidence
that in fact they leave nontrivial imprints on each other and are
better treated as a single entity.
\item The relatively prominent role of the Zeldovich relation argues
for the vacuum structure of QCD to play a central role in shaping
overall vacuum structure. Therefore there is a motivation to
identify candidate "vacuum condensates" specifically associated with
QCD. I like to entertain the notion that there is a condensate of
topological charge N (the quantity canonically conjugate to the
$\theta$ parameter of QCD CP violation). The total mean N in a
Hubble-scale universe is then of order $10^{120}$ instead of the
usually-assumed value of zero. The uncertainty in N from
instanton-induced transitions is then of order $10^{80}$ or so
instead of the usually-assumed value of infinity. And
Zhitnitsky\cite{urban} and his collaborators recently have also
argued for a QCD-related origin of dark energy. Their condensate
does involve vacuum topology, and does connect with the Zeldovich
relation. However, they choose a covariant gauge, and "activate" the
"Veneziano-ghost" degrees of freedom associated with the topological
susceptibility. I find all this very interesting, but would prefer
seeing their arguments recast in a more physical gauge (e.g.
temporal gauge) in the light of my comments in item 3 above.
\end{enumerate}

\section{Acknowledgements}

Thanks go to my colleagues at SLAC and Stanford for patient
criticism of these speculations, and in particular to Ronald Adler,
Jeffrey Scargle, Alex Silbergleit, Robert Wagoner, and Marvin
Weinstein. I also thank Alan Kostelecky' for organizing this superb
meeting, and to Nikodem Poplawski for conversations regarding the
work in reference 18.


\noindent
\begin{thebibliography}{25}
\bibitem{arxiv} For a more general personal overview, see "The Future and its Alternatives",
www.slac.stanford.edu/grp/th/symposia.
\bibitem{bj} J. Bjorken, Ann. Phys. {\bf{24}}, 174 (1963)
\bibitem{NJL} Y. Nambu and G. Jona-Lasinio,Phys. Rev. {\bf{122}}, 345
(1961); {\bf{124}}, 246 (1961)
\bibitem{jdb} J. Bjorken, arXiv:0111196.
\bibitem{nambu}Y. Nambu, Second Meeting on CPT and Lorentz Symmetry, ed. A.
Kostelecky', World Scientific (2001), p. 3.
\bibitem{pam} P. Dirac, Proc. Roy. Soc. {\bf{A209}}, 291 (1951).
\bibitem{sakharov}A. Sakharov, Gen. Rel. Grav. {\bf{32}}, 365 (2000); translation from
Dokl. Akad. Nauk. SSSR {\bf{170}}, 70 (1967).
\bibitem{KT}P. Kraus and E. Tomboulis, arXiv:0203221
\bibitem{HT} M. Henneaux and C. Teitelboim, Phys. Lett. {\bf{B222}}, 195
(1989)
\bibitem{AKpotting}A. Kostelecky' and R. Potting, arXiv:0901.0662, and references
therein.
\bibitem{holst} S. Holst, arXiv:0901.0662
\bibitem{CZ} O. Chandia and J. Zannelli, arXiv:9702025 and arXiv:9708138;
H. Nieh and C.N. Yang, Int. J. Mod. Phys. {\bf{A22}}, 5237 (2007),
and references therein.
\bibitem{gamma} F. Barbero, Phys. Rev. {\bf{D51}}, 5498 and 5507 (1995); G. Immirzi,
arXiv:9701052
\bibitem{bojowald} M. Bojowald and R. Das, Phys. Rev. {\bf{D78}}, 064009 (2008); arXiv:0710.5722.
\bibitem{zeld} Ya. Zeldovich, JETP Lett. {\bf{6}}, 316 (1967).
\bibitem{carniero} For example, see S. Carniero, arXiv:0305081; R. Schutzhold, arXiv:0204018;
A. Randono, arXiv:0805.2955; see also references 18 and 25.
\bibitem{AKetal}A. Kostelecky', N. Russell, and J. Tasson, arXiv:0712.4393
\bibitem{nikodem} N. Poplawski, arXiv:1005.0893; see also A. Randono, arXiv:1005.1294.
\bibitem{levin} J. Levin, Phys. Rept. {\bf{365}}, 251 (2002); arXiv:0108043.
\bibitem{MandM} S. MacDowell and F. Mansouri, Phys. Rev Lett. {\bf{38}}, 739 and 1376
(1977).
\bibitem{FS} L. Freidel and A. Starodubtsev, arXiv:0501191
\bibitem{lisi} See, for example, G. Trayling, arXiv:9912231; G. Lisi, arXiv:0711.0770.
\bibitem{Kodama} H. Kodama, Phys. Rev. D42, 2548 (1990); L. Smolin, hep-th/0209079.
\bibitem{Jackiw} E. Witten, arXiv:0306083 and references therein.
\bibitem{urban} F. Urban and A. Zhitnitsky, arXiv:0906.2162,
arXiv:0906.2165), and arXiv:0909.2684.
\end{thebibliography}
\end{document}